\documentclass{article}

\usepackage{epsfig}
\usepackage{graphicx}
\usepackage{array}
\usepackage{color}
\usepackage{cite}

\def\kzeros        {K$^{0}_{s}$}

\def\lambdazero    {$\Lambda$}

\def\Pgp{\ifmmode{\mathrm \pi}
          \else${\mathrm \pi }$\fi}
\def\Pgpm{\ifmmode{\mathrm \pi^-}
           \else${\mathrm \pi^-}$\fi}
\def\Pgpp{\ifmmode{\mathrm \pi^+}
           \else${\mathrm \pi^+}$\fi}

\def\PK{\ifmmode{\mathrm K}
         \else${\mathrm K}$\fi}
\def\PKpm{\ifmmode{\mathrm K^{\pm}}
           \else${\mathrm K^{\pm}}$\fi}
\def\PKp{\ifmmode{\mathrm K^+}
          \else${\mathrm K^+}$\fi}
\def\PKm{\ifmmode{\mathrm K^-}
          \else${\mathrm K^-}$\fi}
\def\Pp{\ifmmode{\mathrm p}
         \else${\mathrm p}$\fi}
\def\PgL{\ifmmode{\mathrm \Lambda}
          \else${\mathrm \Lambda}$\fi}
\def\PagL{\ifmmode{\overline{\mathrm \Lambda}}
          \else${\overline{\mathrm \Lambda}}$\fi}
\def\PgOm{\ifmmode{\mathrm \Omega^-}
           \else${\mathrm \Omega^-}$\fi}

\def\munit{\ifmmode{\,\mathrm{MeV/{\mit c}^{\,2}}}
          \else{MeV/$c^{\,2}$}\fi}
\def\mup{\ifmmode{\mathrm{\,MeV/{\mit c}}}
          \else{MeV/{\it c}}\fi}
\def\mupp{\ifmmode{\mathrm{\,(MeV/{\mit c})^2}}
          \else{(MeV/{\it c})$^2$}\fi}

\def\gunit{\ifmmode{\,\mathrm{GeV/{\mit c}^{\,2}}}
          \else{GeV/$c^{\,2}$}\fi}
\def\pup{\ifmmode{\mathrm{\,GeV/{\mit c}}}
          \else{GeV/{\it c}}\fi}
\def\pupp{\ifmmode{\mathrm{\,(GeV/{\mit c})^2}}
          \else{(GeV/{\it c})$^2$}\fi}
\def\pum{\ifmmode{\mathrm{\,(GeV/{\mit c})^{-1}}}
          \else{(GeV/{\it c})$^{-1}$}\fi}
\def\pumm{\ifmmode{\mathrm{\,(GeV/{\mit c})^{-2}}}
          \else{(GeV/{\it c})$^{-2}$}\fi}

\def\rmunit{\ifmmode{\,\mathrm{íÜ÷/c^2}}
          \else{íÜ÷/c$^2$}\fi}
\def\rmup{\ifmmode{\mathrm{\,íÜ÷/c}}
          \else{íÜ÷/c}\fi}
\def\rmupp{\ifmmode{\mathrm{\,(íÜ÷/c)^2}}
          \else{(íÜ÷/c)$^2$}\fi}

\def\rgunit{\ifmmode{\,\mathrm{çÜ÷/c^2}}
          \else{çÜ÷/c$^2$}\fi}
\def\rpup{\ifmmode{\mathrm{\,çÜ÷/c}}
          \else{çÜ÷/c}\fi}
\def\rpupp{\ifmmode{\mathrm{\,(çÜ÷/c)^2}}
          \else{(çÜ÷/c)$^2$}\fi}
\def\rpum{\ifmmode{\mathrm{\,(çÜ÷/c)^{-1}}}
          \else{(çÜ÷/c)$^{-1}$}\fi}
\def\rpumm{\ifmmode{\mathrm{\,(çÜ÷/c)^{-2}}}
          \else{(çÜ÷/c)$^{-2}$}\fi}

\def\cpc#1#2#3  {{\rm Computer Phys. Comm.}  {\bf#1}, (#3) #2}
\def\npb#1#2#3  {{\rm Nucl. Phys. B}         {\bf#1}, (#3) #2}
\def\plb#1#2#3  {{\rm Phys. Lett. B}         {\bf#1}, (#3) #2}
\def\prd#1#2#3  {{\rm Phys. Rev. D}          {\bf#1}, (#3) #2}
\def\prl#1#2#3  {{\rm Phys. Rev. Lett.}      {\bf#1}, (#3) #2}
\def\sjnp#1#2#3 {{\rm Sov. J. Nucl. Phys.}   {\bf#1}, (#3) #2}
\def\spjl#1#2#3 {{\rm Sov. JETP Lett.}       {\bf#1}, #2 (#3)}
\def\zpc#1#2#3  {{\rm Z. Phys. C}            {\bf#1}, (#3) #2}

\begin{document}
\pagestyle{empty}

\begin{flushright}
hep-ex/02 \\*[8mm]
\today
\end{flushright}
\vspace{1.8cm}

\centerline{
\parbox{12cm}{\bf
\begin{center}
\Large{The Charm Pentaquark and Kinematic Reflection.}
\end{center}
}}
\vspace*{1.7cm}

\begin{center}
\large{
M.Zavertyaev \\[0.5cm]
P.N.Lebedev Physical Institute,117924 Moscow B-333, Russia}
\end{center}
\vspace*{1.7cm}
\hrule
\vspace*{0.5cm}

\noindent{\bf Abstract} \\
A simple example of the kinematic reflection in the $D^*p$ system
at about 3.1 \gunit\ is given.
\vspace*{0.5cm}
\hrule

\vfill

\clearpage

\section{Introduction}
\label{intro}

The anti-charm pentaquark $\theta_c=uudd\overline{c}$ was observed
by the H1 experiment at HERA \cite{h1} in the $D^{*-}p$ and 
$D^{*+}\overline{p}$
modes in $ep$ collision at $\sqrt{s}$ of 300 and 320 GeV. A narrow peak of
50 events was observed at about 3.1 \gunit\ with the width consistent with the
detector resolution. The search for $\theta_c$ was performed by other
experiments \cite{aleph,belle,bes,cdf,focus,zeus} with the negative result.

But irrelevant to the result of other experiments the invariant mass spectrum
of $D^{*-}p$ and $D^{*+}\overline{p}$ presented by  H1 experiment has
the peak and the question is what might be the explanation for this
particular case.

\section {Monte Carlo simulation.}
\label{mc}

The $D^{*+}$ meson was observed via the decay chain 
$D^{*+} \rightarrow D^0\pi^+ \rightarrow K^-\pi^+\pi^+$. For the pentaquark
search one more negative particle (anti-proton in this case) is needed.
Let us try for the moment calculate the invariant mass of three
particles combination  $K^-\pi^+\pi^+$ - decay products of $D^{*+}$ plus
the same $K^-$ but assigning to it the mass of the proton.
In simulation the $D^{*+}$ momentum was uniformly distributed in the 
range of 20-30 \gunit .

\begin{figure*} 
\addtolength{\abovecaptionskip}{10pt}
\centering
\includegraphics[width=15cm]{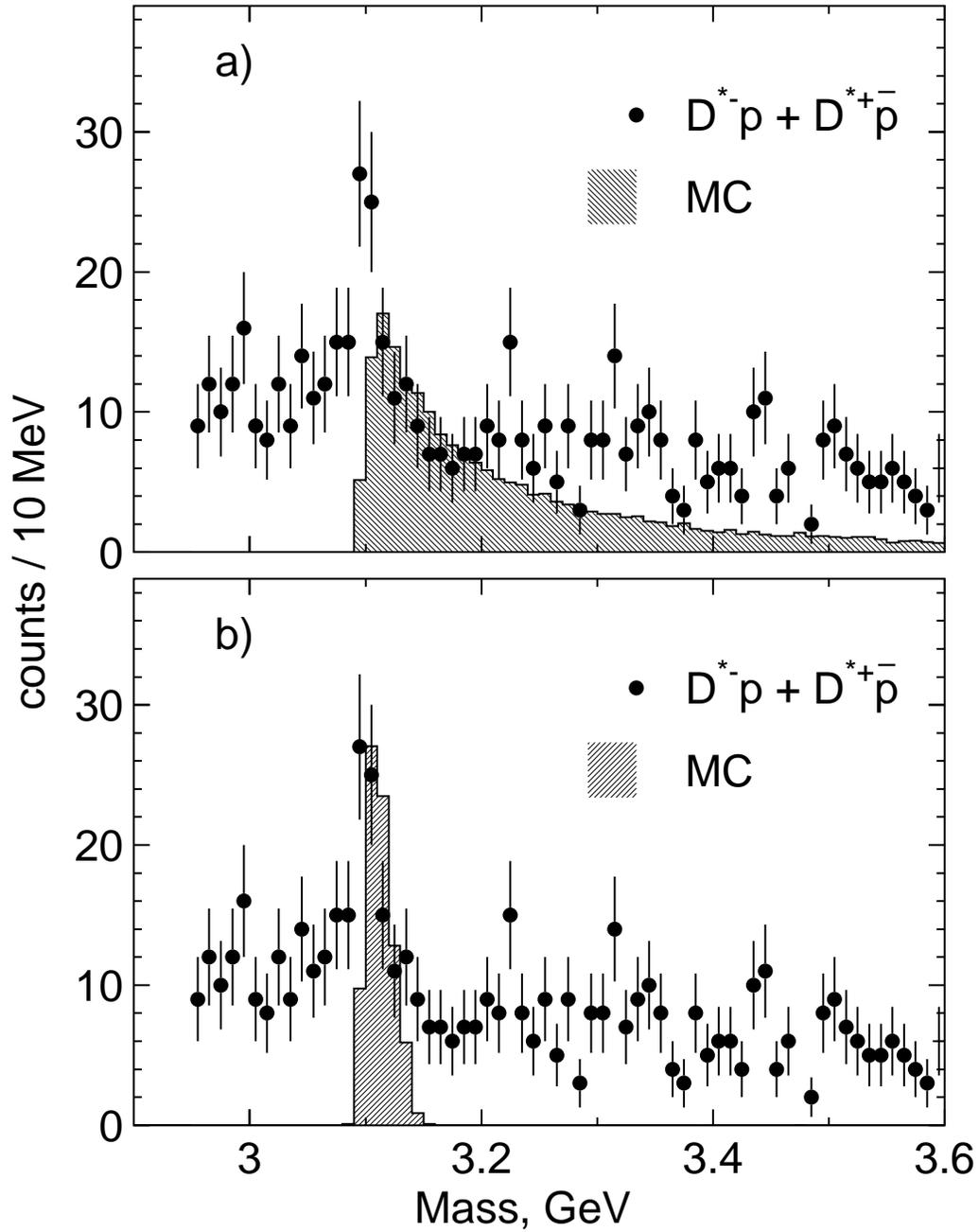}
\caption{ The invariant mass spectra of the 
          $D^{*-}p + D^{*+}\overline{p}$ (taken from \protect\cite{h1})
          and Monte-Carlo simulation: a) no cut on the ``proton'' momentum;
          b) with the cut on the ``proton'' momentum $>$20 \pup.
	 }
\label{fig:h1}
\end{figure*}

The result is presented in Fig.\,\ref{fig:h1} overlapped with
the plot published by H1 experiment \cite{h1}. On the top frame the invariant 
mass of combinations with all ``proton'' momentum is shown, on the bottom
only combinations with the ``proton'' momentum (and kaon as well) above 
20 \pup\ are shown.

The features of the simulated mass spectrum are very interesting. The maximum
of the distribution miraculously coincides with the position of the
experimental pentaquark peak at 3.1 \gunit . The low edge of the
distribution is a kinematic limit for the combination, while the tail at 
higher masses shrinks with the increase of the cut on the negative
particle momentum, so a narrow peak might be created easily.

The same procedure may be applied to $D^{*-}$ with the corresponding changes
and the result of the simulation is identical to $D^{*+}$.

No one should expect that such a primitive Monte-Carlo may reproduce all
the features of the peak in the invariant mass spectrum. The final shape
is a sophisticated convolution of the momentum spectra and mass resolution
of the spectrometer and such comprehensive simulation is not the goal of this 
study. But the principal result is that any ghost track associated with
the kaon in the $D^{*}$ decay may be the reason for the narrow peak generation
at 3.1 \gunit . 

\begin{figure*} 
\addtolength{\abovecaptionskip}{10pt}
\centering
\includegraphics[width=15cm]{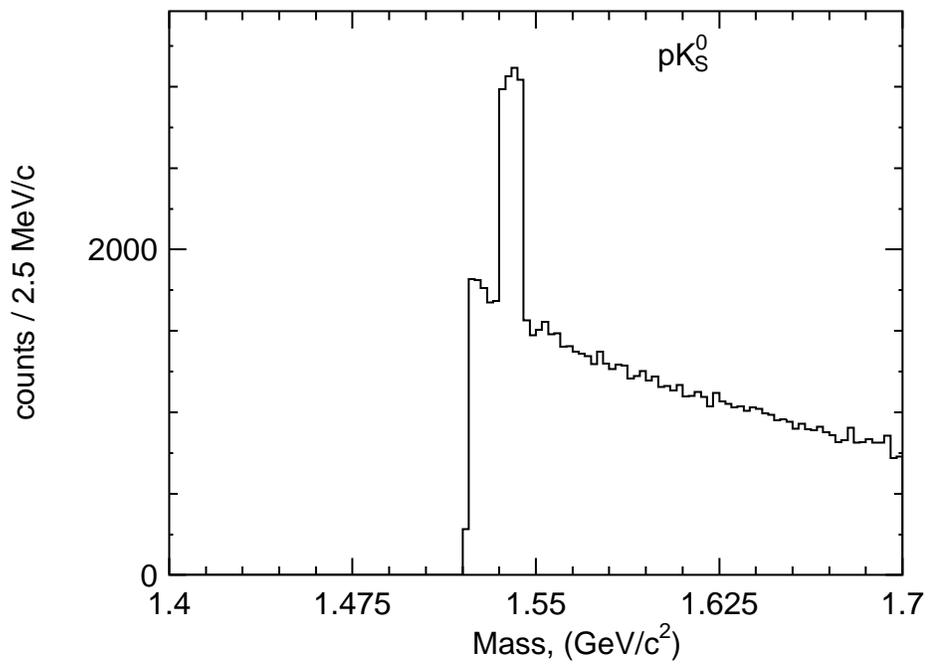}
\caption{ The invariant mass spectrum for p\kzeros\ combinations
          in case of double counting of the proton from \lambdazero\
          decay.
	 }
\label{fig:pk}
\end{figure*}

The problem of double counting or ghost tracks is known from strange 
pentaquark studies \cite{zav,longo} , when the proton from the \lambdazero\ 
used twice creates a sharp peak in the invariant mass distribution exactly at 
1.54\gunit (Fig.\,\ref{fig:pk} ).

\section{Conclusion.}
\label{concl}

A very curious situation exist now in the pentaquark business. For each
channel there is an example of kinematic reflection  which 
coincides with the peak : for $nK$ - see \cite{kinrefl},
for p\kzeros\ - see \cite{zav} and now for $\theta_c$.
Something is really hidden behind this tendency, it
may not be a mere coincidence.

At the end author would like to attract the
attention of the reader to newly published review of the pentaquark
studies \cite{kinref} with many interesting details.

\clearpage

\newpage
\begin{thebibliography}{}

\bibitem{h1}
A.Aktas {\it et al} (H1 Collaboration), 
Phys.Lett. {\bf B588} (2004) 17.

\bibitem{aleph}
S.Schael {\it et al} (ALEPH Collaboration), 
Phys.Lett. {\bf B599} (2004) 1.

\bibitem{belle}
R.Mizuk (BELLE Collaboration), hep-ex/0411005 (23 Nov 2004).

\bibitem{bes}
J.Bei {\it et al}(BES Collaboration),
Phys.Rev. {\bf D70} (2004) 012004.

\bibitem{cdf}
I.Gorelov (CDF Collaboration), hep-ex/0408025 (10 Aug 2004);
D.Litvintsev (CDF Collaboration), \\ hep-ex/0410024 (25 Oct 2004).

\bibitem{focus}
K.Stenson (FOCUS Collaboration), hep-ex/0412021 (7 Dec 2004)

\bibitem{zeus}
S.Chekanov {\it et al} (ZEUS Collaboration),
Phys.Lett. {\bf B591} (2004) 7.

\bibitem{leps}
T.Nakano {\it et al., Evidence for a Narrow S=+1 Baryon resonance
          in photo-production from the Neutron}, 
	  Phys. Rev. Lett. {\bf 91} (2003) 012002-1.

\bibitem{zav}
M.Zavertyaev, hep-ph/0311250

\bibitem{longo}
M.Longo, 2004 Presentation of HyperCP at QN2004 www.qnp2004.org
\bibitem{kinrefl}
A.R.Dziebra {\it et al.},  Phys. Rev. {\bf D69} (2004) 051901.

\bibitem{kinref}
A.R.Dziebra {\it et al.}, hep-ex/0412077 (29 Dec 2004).

\end{thebibliography}
\end{document}